\documentclass[a4paper]{article}

\usepackage{INTERSPEECH2020, amsmath, amssymb, subcaption,cite}

\DeclareMathOperator*{\argmax}{arg\,max}

\newcommand{\refeq}[1]{Eq.~(\ref{eq:#1})}
\newcommand{\reftb}[1]{Table \ref{tb:#1}}
\newcommand{\reffig}[1]{Figure \ref{fig:#1}}
\newcommand{\refsec}[1]{Section \ref{sec:#1}}

\title{Insertion-Based Modeling for End-to-End Automatic Speech Recognition}
\name{Yuya Fujita$^1$, Shinji Watanabe$^2$
    Motoi Omachi$^1$, Xuankai Chang$^2$
    }
\address{
  $^1$Yahoo Japan Corporation, Tokyo, JAPAN\\
  $^2$Center for Language and Speech Processing, Johns Hopkins University, Baltimore, MD, US}
\email{yuyfujit@yahoo-corp.jp}

\begin{document}

\maketitle

\begin{abstract}
End-to-end (E2E) models have gained attention in the research field of automatic speech recognition (ASR). Many E2E models proposed so far assume left-to-right autoregressive generation of an output token sequence except for connectionist temporal classification (CTC) and its variants. 
%Decoding is also performed in the same way. 
However, left-to-right decoding cannot consider the future output context, and it is not always optimal for ASR. 
One of the non-left-to-right models is known as non-autoregressive Transformer (NAT) and has been intensively investigated in the area of neural machine translation (NMT) research. 
One NAT model, mask-predict, has been applied to ASR but the model needs some heuristics or additional component to estimate the length of the output token sequence. 
This paper proposes to apply another type of NAT called insertion-based models, that were originally proposed for NMT, to ASR tasks.
Insertion-based models solve the above mask-predict issues and can generate an arbitrary generation order of an output sequence. 
In addition, we introduce a new formulation of joint training of the insertion-based models and CTC. 
This formulation reinforces CTC by making it dependent on insertion-based token generation in a non-autoregressive manner. 
We conducted experiments on three public benchmarks and achieved competitive performance to strong autoregressive Transformer with a similar decoding condition.
\end{abstract}

\section{Introduction}
\label{sec:intro}
End-to-end (E2E) models have become mainstream in the research field of automatic speech recognition (ASR).
One advantage of the E2E models is the simplicity of the model structure. 
A single neural network receives an acoustic feature sequence and directly generates an output token sequence. It does not need separate models such as acoustic, language and lexicon models commonly used in the conventional ASR system.
There has been a lot of work that aims to improve E2E models \cite{AttentionNIPS2015,LAS2016,amodei2016deep,Prabhavalkar2017,Shinji2017hybrid,Zeyer2018}. 
Several reported that the E2E model achieved comparable or better performance to the conventional ASR system in the product system \cite{Chiu2018google, sainath2020streaming} and publicly available corpora \cite{Park2019,Luscher2019,karita2019comparative} such as Librispeech \cite{Libri2015}, Switchboard \cite{SB1992}, and Corpus of Spontaneous Japanese(CSJ) \cite{CSJ2000}.
%by their in-house data \cite{Chiu2018google, sainath2020streaming}.
%Google researchers reported their E2E model has achieved superior performance to the conventional ASR system not only on the server-side but also on device \cite{Chiu2018google, sainath2020streaming} by their in-house data. 
%Also, the performance of E2E models on publicly available corpora such as Librispeech \cite{Libri2015} and Switchboard \cite{SB1992} has achieved comparable performance to the conventional system \cite{Park2019}.
One of the state-of-the-art E2E models is Transformer \cite{TransformerNIPS2017}, which significantly outperformed the RNN-based E2E models \cite{Karita2019,karita2019comparative}.

Many of the above E2E models assume left-to-right autoregressive generation of an output token sequence. In the speech production context, this assumption is reasonable because speech is produced in a left-to-right order given linguistic content. But in the speech perception context, it is unclear that left-to-right decoding is always the best way. For example, when we listen to speech and encounter a word whose pronunciation is unclear, we leave it as uncertain and re-estimate it using the future context. Mimicking this perceptual process in ASR is scientifically quite important and also has some potential to improve the performance of left-to-right decoding.

One of the non-left-to-right E2E models is known as non-autoregressive Transformer (NAT). It is heavily investigated in the area of neural machine translation (NMT)  \cite{gu2017non, stern2019insertion, Gu2019leven, chan2019kermit}. 
Mask-predict, one of the NAT models, has been applied to speech recognition \cite{Chen2020Listen}. 
It realizes non-autoregressive output token generation by introducing a special token which masks part of an output token sequence. When training the model, some tokens in the output sequence are randomly masked and the model is trained to estimate the masked tokens with non-causal masking in the self-attention of the decoder. 
During decoding, the output token sequence is generated by estimating masked tokens. 
%Note that at the first decoding step, the decoder is fed sequence composed of only special mask tokens. 
There is no causal masking hence non-left-to-right non-autoregressive output sequence generation is realized. 
It achieved competitive performance to an autoregressive model with faster decoding time on AISHELL \cite{aishell_2017}.

However, mask-predict needs some additional component or heuristics to estimate the output token sequence length. 
To overcome this disadvantage, the insertion-based model is proposed \cite{stern2019insertion}. In theory this model can generate an output token sequence with an arbitrary order without any additional component or heuristics to estimate the output token sequence length. In NMT, performance competitive to autoregressive Transformer is reported with fewer iterations in decoding \cite{chan2019kermit}.

This paper proposes using the insertion-based models for E2E ASR with an in-depth investigation of three insertion-based models originally proposed in NMT.
%In this work, in order to confirm E2E ASR with insertion based models works or not, we applied three major insertion based models originally proposed in NMT to ASR task. 
In addition, we introduce a new formulation of 
joint modeling of connectionist temporal classification (CTC) \cite{Graves2006CTC} and insertion-based models.
This formulation can be viewed as modeling the joint distribution between the CTC probability and the insertion-based sequence generation probability. 
Hence the CTC probability depends on the insertion-based output token sequence generation.
With this new formulation, the monotonic alignment property of CTC is reinforced by insertion based token generation. It achieves performance competitive with an autoregressive left-to-right model decoded with a similar decoding condition in non-left-to-right and non-autoregressive manner.

The source code will be publicly available in the open source E2E modeling toolkit ESPnet \cite{Watanabe2018Espnet}. 

\section{Related work}
To the best of our knowledge, this is the first work to apply insertion-based models to ASR tasks. 

As mentioned in \refsec{intro}, this work is another type of NAT application to ASR tasks compared with \cite{Chen2020Listen}. 
Our work does not use a special mask token or need to estimate an output sequence length in advance. Furthermore, our work can handle autoregressive and non-autoregressive models in a single formulation and also introduces a new formulation of joint modeling of CTC and insertion-based models.

Non-autoregressive E2E ASR using a CTC-like model is proposed in Imputer \cite{chan2020imputer}. It assumes that the alignment at the $i$-th generation step depends on the past $(i-1)$-th alignment. The alignment is estimated similarly to mask-predict in a non-autoregressive manner. Our work is different in using insertion-based models and joint modeling of CTC and insertion-based token sequence generation.

%In \cite{Kubo2020joint}, joint distribution modeling between two different sequences is proposed but their model is not non-left-to-right nor non-autoregressive. 
%phoneme and grapheme sequence is proposed. 
%Our work is different in using an insertion based model for joint distribution modeling. 
%In addition, joint modeling with CTC probability is not used in their work.

\section{Insertion-based end-to-end models}
\label{sec:method}
%The end-to-end (E2E) model directly calculates posterior probabilities over the output token sequence by a single neural network. 
Let $X=\{\mathbf{x}_t \in \mathbb{R} ^{d} | t=1,\cdots,T\}$ be a $d$-dimensional acoustic feature sequence and $C=\{c_n \in \mathcal{V} | n=1,\cdots,N \}$ be an output token sequence.
$T$ is the input length, $\mathcal{V}$ is a set of distinct tokens, and $N$ is the output length. 
Then, decoding of the E2E model is performed to maximize the posterior probability $p^{\text{e2e}}(C|X)$:
\begin{equation}
\label{eq:e2e}
    \hat{C} = \argmax _{C} p^{\text{e2e}}(C|X).
\end{equation} 
Training of the E2E model is also based on this criterion. 
The difference between various E2E models is how to define the posterior $p^{\text{e2e}}(C|X)$ in \refeq{e2e}. 

In the insertion-based models, it is assumed to be marginalized over all possible insertion orders (permutation). Let $Z = \{z_n \in \mathbb{N}^1 | n=1, \cdots, N \}$ be an insertion order.
For example, suppose $C=\{\text{this, is, a, pen}\}$, $Z$ is all the permutations of ordering 4 tokens, i.e., 
\begin{align}
 & Z \in \{\{4,1,2,3\}, \{4,2,1,3\}, \cdots, \{1,2,3,4\} \},
\end{align}
\begin{equation}
\left\{
\begin{alignedat}{4} 
& & C^{Z=\{4,1,2,3\}}&=\{\text{pen, this, is, a}\} \\
& & C^{Z=\{4,2,1,3\}}&=\{\text{pen, is, this, a}\}  \\
& & & \vdots \\
& & C^{Z=\{1,2,3,4\}}&=\{\text{this, is, a, pen}\}
\end{alignedat}
\right. ,
\end{equation}
where $C^{Z} = \{c_{z_n} \in \mathcal{V} | n=1,\cdots,N \}$ is the permutated output token sequence with an insertion order $Z$.
The number of all permutations $|Z|$ is ${}_4 P _4$.
Then, the posterior in \refeq{e2e} is factorized with the sum and product rules as:
\begin{align}
\label{eq:ins_model}
    p^{\text{e2e}}(C|X) & = \sum _Z p(C, Z|X) = \sum _Z p(C^Z |X) p(Z). 
                        %& = \sum _Z p(C|X,Z) p(Z|X), \\
                        %& = \sum _Z p(C|X,Z) p(Z).
\end{align}
We assume that insertion order $Z$ does not depend on input feature hence $P(Z|X) = P(Z)$.

Definition of $p(C^Z |X)$ in \refeq{ins_model} is different between insertion-based models and explained in the next subsection.
Note that the left-to-right autoregressive model can be interpreted as a special case where $p(Z)$ is fixed to be the left-to-right order $p(Z = \{1,2,\cdots,N\}) = 1$ and $p(C^Z |X)$ in \refeq{ins_model} is $p(C^Z |X) =: \prod _{i=1} ^N p(c_n|c_{1:n-1}, X)$. 
%\begin{align}
    %& p(Z = \{1,2,\cdots,N\}) = 1, \\
%    & p(C|X,Z) =: \prod _{i=1} ^N p(c_n|c_{1:n-1}, X).
%\end{align}

When training the model of \refeq{ins_model}, a lower bound of log-likelihood $\mathcal{L}(\theta)$ where $\theta$ is the parameters of the model is maximized under a predefined prior distribution over insertion order $Z$. 
%\begin{align}
%    \log p^{\text{e2e}}(C|X) & = \log \sum _Z p(C|X,Z) p(Z), \\
%                            & \geq \sum _Z p(Z) \log p(C|X,Z) =: \mathcal{L}(\theta)
%\end{align}
\begin{align}
    \log p^{\text{e2e}}(C|X) & \geq \sum _Z p(Z) \log p(C^Z|X) =: \mathcal{L}(\theta)
\end{align}
From the next subsection, three existing insertion-based models are explained. 

\subsection{Insertion-based decoding}
Insertion-based decoding (InDIGO) \cite{gu2019insertion} is an insertion-based model using Transformer with relative position representation. 
%Let $C^{Z} = \{c_{z_n} \in \mathcal{V} | n=1,\cdots,N \}$ be a permutated output token sequence with insertion order $Z$ and 
Let $\mathbf{R}^Z_n \in \{-1,0,+1\}^{n \times n}$ be a relative position representation at generation step $n$ under an insertion order $Z$.
Element $r^Z_{ij}$ is defined as:
\begin{equation}
    r^Z_{ij} = \left\{ 
    \begin{alignedat}{3}
             -1 & \quad & z_j > z_i \\
             0  & & z_j = z_i \\
             1  & & z_j < z_i
    \end{alignedat}
    \right. .
\end{equation}
$p(C^Z|X)$ in \refeq{ins_model} of InDIGO is defined as:
\begin{align}
        p(C^Z|X) & =: p(C^Z, \mathbf{R}^Z_N|X) \\
     & = \prod _{n=1} ^N p(c^Z_{n}, \mathbf{r}^Z_n | c^Z_{1:n-1}, \mathbf{R}^Z_{n-1}, X),
    \label{eq:indigo_pst}
\end{align}
where $\mathbf{r}^Z_n$ is the $n$-th column vector of $\mathbf{R}^Z_n$.
The factorized form $p(c^Z_{n}, \mathbf{r}^Z_n | c^Z_{1:n-1}, \mathbf{R}^Z_{1:n-1}, X)$ in \refeq{indigo_pst} is modeled by Transformer. Let $\mathbf{H}_{\text{dec}} \in \mathbb{R}^{b \times N}$ be the final output of the decoder layer of Transformer where $b$ is the dimension of self-attention, 
\begin{align}
      p(c^Z_{n}, \mathbf{r}^Z_n | c^Z_{1:n-1}, \mathbf{R}^Z_{1:n-1}, X)
    =: \underbrace{p(c^Z_{n}|\mathbf{H}_{\text{dec}})}_{\text{Word prediction}}  \underbrace{p(\mathbf{r}_n|c^Z_{n}, \mathbf{H}_{\text{dec}})}_{\text{Position prediction}}.
    \label{eq:indigo_final}
\end{align}
For the word prediction term, a linear transform and softmax operation are applied to $\mathbf{H}_{\text{dec}}$, and for the position prediction term a pointer network \cite{Vinyals2015pointer} is used.

During decoding, the next token to be inserted is estimated by the word prediction term then its position is estimated by the position prediction term in \refeq{indigo_final}.
Because of this sequential operation, only a single token is generated per iteration during decoding. Therefore, InDIGO can be non-left-to-right but requires $N$ iterations.
%Note that the internal state of decoder for past tokens do not change by one iteration of decoding step, InDIGO can be non-left-to-right and autoregressive (Only single token is generated per iteration).

\subsection{Insertion Transformer and KERMIT}
\begin{figure}[t]
  \centering
    \begin{tabular}{c}
        \begin{minipage}[t]{6cm}
            \centering
          \includegraphics[width=6cm]{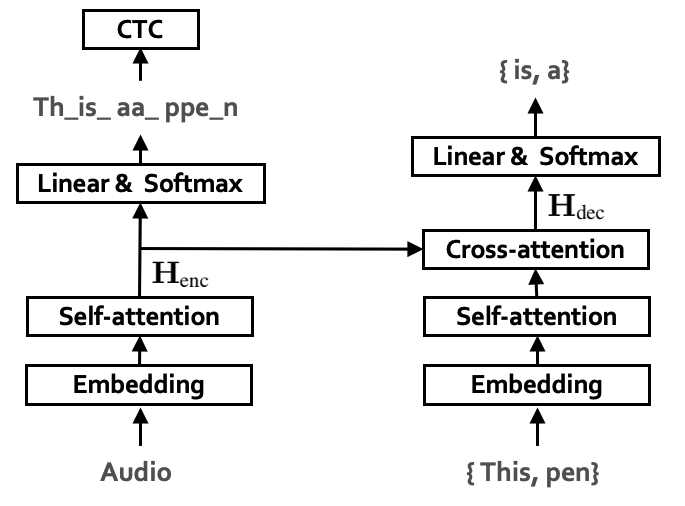}
            \subcaption{Insertion Transformer.}\label{fig:inst}
        \end{minipage}  \\
        \begin{minipage}[t]{6cm}
        \centering
         \includegraphics[width=6cm]{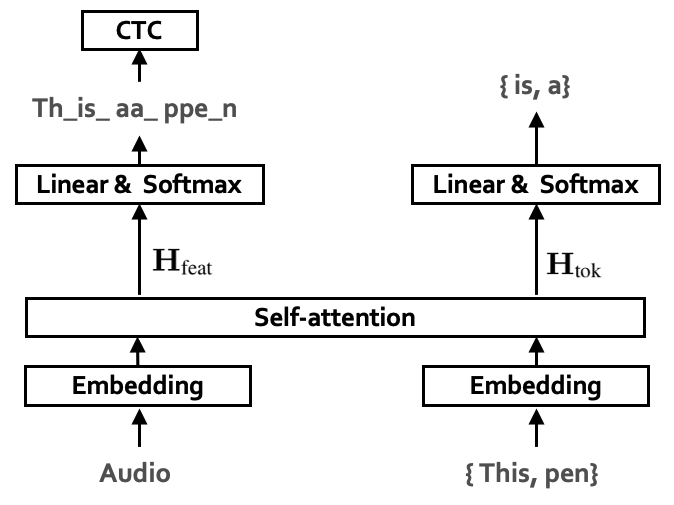}
        \subcaption{KERMIT.}\label{fig:kermit}
        \end{minipage}
    \end{tabular}
    \caption{Schematic diagram of hybrid training with CTC. In the case of KERMIT, CTC is dependent on both audio and output token.}
    \label{fig:hybrid}
\end{figure}

Another type of insertion-based model is Insertion Transformer \cite{stern2019insertion} and KERMIT (Kontextuell Encoder Representations Made by Insertion Transformations) \cite{chan2019kermit}. The basic formulation of these two models is the same. Let $c^Z_n$ be a token to be inserted and $l^Z_n$ be a position where the token is inserted at the $n$-th generation step under an insertion order $Z$.  
$p(C^Z|X)$ in \refeq{ins_model} is defined as:
\begin{align}
            p(C^Z|X) 
        =: & \prod _{n=1} ^N p \left( 
            \left( c^Z_n, l^Z_n \right)
                | \left( c^Z_1, l^Z_1 \right), \cdots, \left( c^Z_{n-1}, l^Z_{n-1} \right)
                ,X \right) \nonumber \\
        =  & \prod _{n=1} ^N p\left( \left( c^Z_n, l^Z_n \right) | {c^{Z}}^{\prime}_{1:n-1}, X \right),
        \label{eq:inst_model}
\end{align}
where ${c^{Z}}^{\prime}_{1:n-1}$ is the sorted token sequence at the $n$-th generation step.
For example, in case of $C=\{\text{this, is, a, pen}\}$ and $Z = \{3, 1, 4, 2\}$, i.e., $C^Z=\{\text{a, this, pen, is}\}$,
\begin{align}
    &n=1: (c^Z_1, l^Z_1) = (\text{a},1), & & {c^Z}^{\prime}_{1} &= & \{\text{a}\},\nonumber \\ 
    &n=2: (c^Z_2, l^Z_2) = (\text{this},1), & & {c^Z}^{\prime}_{1:2} &= & \{\text{this, a}\},\nonumber \\
    &n=3: (c^Z_3, l^Z_3) = (\text{pen},3), & & {c^Z}^{\prime}_{1:3} &= & \{\text{this, a, pen}\},\nonumber \\
    &n=4: (c^Z_4, l^Z_4) = (\text{is},2), & & {c^Z}^{\prime}_{1:4} &= & \{\text{this, is, a, pen}\}. \nonumber
\end{align}
Note that $l_n^Z$ is a position relative to the hypothesis at the $(n-1)$-th generation step. In the case of $n=3$ explained above, $l_n^Z \in \{1,2\}$ because there are two tokens in the previous hypothesis.
%\begin{description}
%    \item[$n=1$:] $(c^Z_1, l^Z_1) = (\text{a},1), {C^Z}^{\prime}_{1} = \{\text{a}\}$, 
%    \item[$n=2$:] $(c^Z_2, l^Z_2) = (\text{this},1), {C^Z}^{\prime}_{1:2} = \{\text{this, a}\}$,
%    \item[$n=3$:] $(c^Z_3, l^Z_3) = (\text{pen},2), {C^Z}^{\prime}_{1:3} = \{\text{this, a, pen}\}$,
%    \item[$n=4$:] $(c^Z_1, l^Z_1) = (\text{is},1), {C^Z}^{\prime}_{1:4} = \{\text{this, is, a, pen}\}$.
%\end{description}

The difference between Insertion Transformer and KERMIT is the matrix $\mathbf{H} \in \mathbb{R}^{b \times N} $ used when the posterior is calculated.
In the case of Insertion Transformer, the final output of the decoder layer of Transformer $\mathbf{H}_{\text{dec}} \in \mathbb{R}^{b \times N} $ is used as $\mathbf{H}$. On the other hand, KERMIT uses only the encoder block of Transformer. Acoustic feature sequence and token embedding are concatenated and fed into the encoder block.
The final output of the encoder layer is sliced as $\mathbf{H}_{\text{tok}} \in \mathbb{R}^{b \times N} $ then used as $\mathbf{H}$. This difference is depicted in \reffig{hybrid}.

By using the $\mathbf{H}$ matrix, the posterior in \refeq{inst_model} is calculated as:
\begin{align}
      p\left( \left( c^Z_n, l^Z_n \right) | {c^Z}^{\prime}_{1:n-1}, X \right)
    =: \underbrace{p(c^Z_n | l^Z_n, \mathbf{H})}_{\text{Word prediction}} 
        \underbrace{p(l^Z_n | \mathbf{H})}_{\text{Position prediction}}.
        \label{eq:inst_pst}
\end{align}
The word and position prediction term is calculated by a softmax followed by a linear transformation of $\mathbf{H}$.

There are two ways of decoding. The first one is autoregressive greedy decoding directly using the posterior in \refeq{inst_pst}:
\begin{align}
    (\hat{c}, \hat{l}) = \argmax_{c,l}  p\left( \left( c, l \right) | {c}^{\prime}_{1:n-1}, X \right).
\end{align} 
The second way is non-autoregressive parallel decoding using only the word prediction term in \refeq{inst_pst}:
\begin{align}
    \hat{c} = \argmax _{c}  p\left( c |l, {c}^{\prime}_{1:n-1}, X \right).
\end{align}
When the balanced binary insertion order proposed in \cite{stern2019insertion} is used as $p(Z)$, parallel decoding finishes empirically with $\log_2(N)$ iterations. This order is to insert centermost tokens of current hypothesis. For example, suppose $C=\{c_1, \cdots, c_9\}$, then the hypothesis grows like $\{c_5\} \rightarrow \{c_3, c_5, c_7\} \rightarrow \{c_2, c_3, c_4, c_5, c_6, c_7, c_8\} \rightarrow \{c_1, c_2, c_3, c_4, c_5, c_6, c_7, c_8, c_9\} $.

\section{Insertion-based/CTC joint modeling}
\label{sec:proposed}
Speech is generated in a left-to-right order hence the alignment is monotonic. Therefore, an E2E model trained with a CTC objective is reported to achieve faster convergence and high accuracy \cite{Shinji2017hybrid}. It is natural to apply this technique to insertion-based models. In the case of InDIGO and Insertion Transformer, its network is composed of an encoder and decoder so it is easy to apply this technique the same way as in \cite{Shinji2017hybrid}. However, for KERMIT, because it consists of only an encoder, a new formulation must be introduced.

Let $Y$ be an output token sequence to be modeled by CTC. Usually, $Y$ is set as $Y=C$. Joint modeling is to extend $P(C^Z|X)$ in \refeq{ins_model} as:
\begin{align}
    P(C^Z|X)  =:  P(C^Z, Y|X)  
             =  P(Y|X,C^Z) P(C^Z|X).
            \label{eq:joint_ctc}
\end{align}
The term $P(Y|X,C^Z)$ in \refeq{joint_ctc} is modeled by CTC. Let $A$ be a sequence of tokens extended with a blank symbol, $A=\{a_t \in \mathcal{V} \cup \{\text{blank}\}|t=1,\cdots,T\}$. $\mathcal{F}(\cdot)$ is a mapping function which deletes repetitions and the blank symbol from a sequence $A$ hence $\mathcal{F}(A) = Y$. The CTC probability is formulated as:
\begin{align}
    p(Y|X,C^Z) &=: \sum _{A \in \mathcal{F}^{\text{-1}}(Y)} p(A|X,C^Z).
    \label{eq:ctc}
\end{align}
In the case of InDIGO and Insertion Transformer, the final output of the encoder layer 
$ \mathbf{H}_{\text{enc}} \in \mathbb{R}^{b \times T} $ 
is used to calculate $p(A|X,C^Z)$ in \refeq{ctc} as:
\begin{align}
    p(A|X,C^Z) \simeq p(A|X) =: p(A|\mathbf{H}_{\text{enc}}),
        \label{eq:ctc_approx}
\end{align}
where $p(A|\mathbf{H}_{\text{enc}})$ is calculated by applying a linear transformation and softmax to $\mathbf{H}_{\text{enc}}$.

For KERMIT, $p(A|X,C^Z)$ in \refeq{ctc} can not be approximated as in \refeq{ctc_approx}. KERMIT consists of only an encoder and acoustic feature sequence and token embedding are concatenated then fed into the encoder block. Therefore, the output of the encoder block depends on both the acoustic feature and output token sequence. 
There might be several ways how to calculate $p(A|X,Z,C)$ in \refeq{ctc}. In this work, output of KERMIT encoder is sliced as $\mathbf{H}_{\text{feat}} \in \mathbb{R}^{b \times T}$ and used: 
\begin{align}
    p(A|X,C^Z) &=: p(A|\mathbf{H}_{\text{feat}}).
\end{align}
This process is depicted in \reffig{kermit}.
%The interesting property of this formulation is that the CTC probability depends on the output of insertion-based token generation. Usually, CTC depends on only the acoustic feature sequence hence is completely non-autoregressive. 
This formulation can reinforce CTC by making it dependent not only on the acoustic feature sequence but also on the output token sequence from insertion-based generation. Note that this formulation still retains non-autoregressive characteristics.

When training the model, in order to adjust the range of the two terms in \refeq{joint_ctc}, the CTC weight $\alpha$ is introduced as:
\begin{align}
        & \log P(Y|X,C^Z) P(C^Z|X) \nonumber \\
   \simeq & \alpha \log P(Y|X,C^Z) + (1.0 - \alpha) \log P(C^Z|X).
\end{align}
During decoding, either the CTC part $p(Y|X,C^Z)$ or the insertion part $p(C^Z|X)$ in \refeq{joint_ctc} can be used.

%\begin{table}[t]
%  \caption{Parameters used for baseline and insertion based models.}
%  \label{tb:param}
%  \centering
%  \begin{tabular}{r|rrcr}
%            & \#Encoder & \#Decoder & $p(Z)$ & \#Epochs \\
%    \hline \hline
%        CTC &  18 & - & - & 50 \\
%    \hline
%        AT  &  12 & 6 & - & 50 \\
%    \hline
%        InDIGO & 12 & 6 & L2R & 50 \\
%    \hline
%        Insertion  & 12 & 6 & L2R & 50 \\
%        Transformer  &  &  & BBT & 300 \\
%    \hline
%        KERMIT & 18 & - & L2R & 50 \\
%         &  &  & BBT & 300 
%  \end{tabular}
%\end{table}

\begin{table*}[tb]
  \caption{CER of CSJ, AISHELL and WER of TEDLIUM2.\protect\footnotemark }
  \label{tb:results}
 \begin{center}
  %\small
  %\scalebox{0.85}{
    \begin{tabular}{r|rrc|rr||rrr|rr|rr}
       & & & & \multicolumn{2}{c||}{CTC weight $\alpha$} & \multicolumn{3}{c|}{\textit{\textbf{CSJ 271h}}} & \multicolumn{2}{c|}{\textit{\textbf{TEDLIUM2}}} & \multicolumn{2}{c}{\textit{\textbf{AISHELL}}} \\
        Model & Beam & Iterations & $p(Z)$ & train & decode & \textit{\textbf{Eval1}} & \textit{\textbf{Eval2}} & \textit{\textbf{Eval3}} & \textit{\textbf{dev}} & \textit{\textbf{test}} & \textit{\textbf{dev}} & \textit{\textbf{test}} \\
      \hline \hline
       %AT & 40 & - & 0.3 & 0.3 &   - &   - &    - &  9.0 &  7.9 & -   &   - \\
    %      & 20 & - &     &     & 7.6 & 5.1 & 13.0 &    - &    - & -   &   - \\
      AT  & 10 & $10N$ & - & 0.3 & 0.3 & 7.9 &  5.7 &  13.7 &  10.6 & 9.1 & 6.5 & 7.2 \\
          &  1 & $N$ &   &     &     & 8.1 & \textbf{5.4} & 13.9 & 12.7 & 10.1 & 6.7 & 8.1 \\
      \hline
   InDIGO &  1 & $N$ & L2R & 0.0 & 0.0 & 8.4 & 6.2 & 14.7 & - & - & - & - \\
          &    &  &     & 0.3 &     & \textbf{7.8} & 5.5 & \textbf{13.3} & 13.6 & \textbf{9.6} & \textbf{6.1} & \textbf{6.7}   \\
\hline         
   Insertion &  1 & $N$ & L2R & 0.0 & 0.0 & 8.7 & 6.3 & 16.1 &  - &  - & -  & - \\
 Transformer &    &     &     & 0.3 &     & 8.3 & \textbf{5.4} & 13.9 & \textbf{11.2} &  \textbf{9.6} & 6.8 & 7.6    \\
\hline
   KERMIT &  1 & $N$ & L2R & 0.0 & 0.0 & 11.0 &  8.0 & 18.9 &   -  &    -  &  -  &   -  \\
          &    &     & & 0.9 &     & 9.2  &  6.7 & 15.7 & 14.9 &  12.4 & 7.7 & 8.9\\
          &    &     & &     & 1.0 & 9.5  &  6.7 & 14.8 &  16.1 &  15.4  & 7.8 & 8.8\\
\hline        
\hline
    CTC & 1 & $1$ & - & 1.0 & 1.0 & 8.5 & 6.1 & 13.8 & 16.1 & 16.3 & \textbf{6.8} & \textbf{7.6} \\
\hline
   Insertion & 1 & $\simeq \log_2(N)$  & BBT & 0.0 &  0.0 & 15.0 & 12.4 & 21.6 &  - &  -   & -   &  - \\
Transformer &    & &     & 0.3 &     & 14.1 & 10.8 & 18.0 & 19.1 & 16.3 & 9.6 &  10.6 \\
\hline
   KERMIT &  1  & $\simeq \log_2(N)$ & BBT & 0.0    & 0.0 & 12.5 &  9.7 & 18.5 &  -   &  -    &  -  & -  \\
          &    &    & & 0.9 (Proposed &     & 11.5 &  9.1 & 16.7 & 18.8 &  15.0 & 9.8 & 10.9 \\
          &    &  & &     formulation) & 1.0 &  \underline{\textbf{7.2}} &  \underline{\textbf{5.1}} & \underline{\textbf{12.5}} & \underline{\textbf{10.5}} &  \underline{\textbf{9.8}} & \underline{\textbf{6.7}} & \underline{\textbf{7.5}}
    \end{tabular}
 \end{center}
\end{table*}

\section{Experiments}
\subsection{Setup}
We used three corpora, the Corpus of Spontaneous Japanese (CSJ) \cite{CSJ2000}, TEDLIUM2 \cite{tedlium2} and AISHELL\cite{aishell_2017}. 
As a baseline model we chose CTC \cite{Graves2006CTC}, which is similar to the work in \cite{salazar2019sactc} using the Transformer encoder layers. 
Another baseline is autoregressive Transformer (AT) \cite{Karita2019}. 

The three insertion-based models described in \refsec{method} are compared to the baseline. For parameters with these models, we followed the Transformer recipe of ESPnet \cite{Watanabe2018Espnet} based on \cite{Karita2019}. 
The numbers of layers for the encoder and decoder were 12 and 6, respectively.
We increased the number of encoder layers to 18 for CTC and KERMIT because they are composed of only an encoder.  
%Parameters changed to this experiment are summarized in \reftb{param}.
%We increased the number of encoder layers to 18 for CTC and KERMIT that are composed of only encoder to match the parameter of another model comprised of encoder and decoder.  
We focused on two types of priors, left-to-right (L2R) and balanced binary tree (BBT) for $p(Z)$ in Eq.~\eqref{eq:ins_model} to simplify the comparison. L2R is evaluated in order to see if there is a performance difference to AT from explicit modeling of the insertion position of tokens. BBT is chosen because it can decode an $N$ length sequence with empirically $\log_2(N)$ iterations.
In training with the BBT prior, we increased the number of epochs from 50 to 300 because only a single step of the output token sequence generation is trained in one minibatch while with the L2R prior we can train the whole sequence generation. 

Since our insertion-based models do not have a beam search algorithm, we mainly compare the L2R-prior insertion-based models with AT (beam=1) and the BBT-prior insertion-based models with CTC (beam=1).
%What we expect in this experiment is that the insertion based model with the L2R prior achieves comparable performance to AT without beam search (beam=1) and with the BBT prior it achieves better performance at least to CTC, and competitive performance to AT.

  \footnotetext{The results of proposed formulation has been updated since the last submission because we found a bug in the code.}
  
\subsection{Results}
First, the results of AT and insertion-based models with the L2R prior are shown in the upper part of \reftb{results}. %Autoregressive Transformer with a beam size of greater than one achieved the best performance. But 
When compared to AT without beam search, the performance of insertion-based models trained with the CTC objective is mostly better except for the eval2 set of CSJ. 
%Explicit modeling of insertion position is effective when combined with hybrid training with CTC objective and decoded with a beam size of one.

Next, the models with the BBT prior are compared in the lower part of \reftb{results}. 
These are non-autoregressive models hence performance is first compared to CTC. Unfortunately, Insertion Transformer with the BBT prior cannot compete with CTC even with hybrid training with CTC. 
KERMIT with joint training with CTC, which is a new formulation introduced in \refsec{proposed}, achieved better performance than CTC on all the corpora. 
Notably, it achieved better performance in a non-autoregressive manner than AT without beam search as highlighted by underlined numbers in \reftb{results}. 
Furthermore, on CSJ, the pefromance is better than AT with beam search.
%However, on AISHELL, the performance was not as good as CTC.

\subsection{Discussion}
One of the remarks we got from the experiments of the L2R prior is that explicit modeling of the position of an output token and hybrid training with the CTC objective worked complementarily and the quality of a hypothesis in decoding was improved.
%insertion-based models trained with the L2R prior and CTC objective achieved better performance than AT without beam search except
%This is probably because explicit modeling of the position of an output token and hybrid training with the CTC objective worked complementarily and the quality of a hypothesis in decoding was improved. 

Another remark is that when the BBT prior is used, the new formulation of joint training with CTC introduced in this paper seems to make use of both benefits of left-to-right and non-left-to-right generation orders. 
%can achieve reasonable performance with the new formulation of joint training with CTC introduced in this work. 
%CTC assumes left-to-right and non-autoregressive generation. 
%The new formulation seems to introduce the complementary effect of taking the benefit of left-to-right and non-left-to-right generation orders.
However, the performance improvement of the BBT prior on AISHELL from AT without beam search is smaller than other corpora.
In this case, the performance of CTC is close to that of AT unlike the other tasks, and the effectiveness of the generation order may depend on the language or task.
%suggests this is not always true. In this case, performance of CTC is close to the autoregressive Transformer with a beam size of one. This suggests that left-to-right modeling is enough on this data.

\section{Conclusions}
This paper proposes applying three insertion-based models, originally proposed for NMT, to ASR tasks. In addition, we introduce a new formulation for joint training of the insertion-based model and CTC. 
Our experiments show that InDIGO and Insertion Transformer trained with the L2R prior achieved comparable or better performance than autoregressive Transformer without beam search. 
Models trained with the BBT prior and the proposed formulation, which retains non-autoregressive characteristics, achieved better performance than CTC and competitive with autoregressive Transformer without beam search on CSJ, TEDLIUM2 and AISHELL.
The number of iteration required for decoding is smaller than AT but an investigation of real time factor is left as future work.
%Therefore, we will investigate more solid use of insertion-based models including an extension of decoding algorithm with beam search. 
%However, this effectiveness is not always true and we will investigate more solid use of insertion-based models including an extension of their decoding algorithm with beam search. 

\bibliographystyle{IEEEtran}

\bibliography{mybib}

\end{document}